\documentclass{aastex63}

\usepackage{multirow}
\usepackage{graphicx,epsfig,natbib,color}
\usepackage{epstopdf}
\usepackage{subfigure}
\usepackage{threeparttable}
\usepackage{graphicx}
\usepackage{epsfig}
\usepackage{natbib}
\received{...} %
\revised{...}
\accepted{...}

\submitjournal{ApJ}

\shorttitle{Thermodynamic Evolution of Solar Flare Supra-arcade Downflows}
\shortauthors{Li et al.}

\begin{document}

\title{Thermodynamic Evolution of Solar Flare Supra-arcade Downflows}

\correspondingauthor{Xin Cheng}
\email{xincheng@nju.edu.cn}

\author{Z. F.  Li}
\affiliation{School of Astronomy and Space Science, Nanjing University, Nanjing, 210046, People's Republic of China}
\affil{Key Laboratory of Modern Astronomy and Astrophysics (Nanjing University), Ministry of Education, Nanjing 210093, China\\}
\author{X. Cheng}
\affiliation{School of Astronomy and Space Science, Nanjing University, Nanjing, 210046, People's Republic of China}
\affil{Key Laboratory of Modern Astronomy and Astrophysics (Nanjing University), Ministry of Education, Nanjing 210093, China\\}
\affil{Max Planck Institute for Solar System Research, Gottingen, D-37077, Germany}
\author{M. D. Ding}
\affiliation{School of Astronomy and Space Science, Nanjing University, Nanjing, 210046, People's Republic of China}
\affil{Key Laboratory of Modern Astronomy and Astrophysics (Nanjing University), Ministry of Education, Nanjing 210093, China\\}
\author{Katharine K. Reeves}
\affiliation{Harvard-Smithsonian Center for Astrophysics, 60 Garden St. MS 58, Cambridge, MA 02138, USA}
\author{DeOndre Kittrell}
\affiliation{Morgan State University, Baltimore, MD 21212, USA}
\affil{Harvard-Smithsonian Center for Astrophysics, 60 Garden St. MS 58, Cambridge, MA 02138, USA}
\author{Mark Weber}
\affiliation{Harvard-Smithsonian Center for Astrophysics, 60 Garden St. MS 58, Cambridge, MA 02138, USA}
\author{David E. McKenzie}
\affiliation{NASA Marshall Space Flight Center, 320 Sparkman Dr NW, Huntsville, AL 35805, USA}

%Abstract
\begin{abstract}
Solar flares are rapid energy release phenomena that appear as bright ribbons in the chromosphere and high-temperature loops in the corona, respectively. Supra-arcade Downflows (SADs) are plasma voids that first come out above the flare loops and then move quickly towards the flare loop top during the decay phase of the flare. In our work, we study 20 SADs appearing in three flares. By differential emission measure (DEM) analysis, we calculate the DEM weighted average temperature and emission measure (EM) of the front region and the main body of SADs. It is found that the temperatures of the SAD front and body tend to increase during the course of SADs flowing downwards. The relationship between the pressure and temperature fits well with the adiabatic equation for both the SAD front and body, suggesting that the heating of SADs is mainly caused by adiabatic compression. Moreover, we also estimate the velocities of SADs via the Fourier Local Correlation Tracking (FLCT) method and find that increase of the temperature of the SAD front presents a correlation with the decrease of the SAD kinetic energy, while the SAD body does not, implying that the viscous process may also heat the SAD front in spite of a limited role.
%1. coronal heating or what else?
%
\end{abstract}

%Introduction
\section{Introduction}\label{intro}

Solar flares are one of the most energetic phenomena in the solar system. Supra-arcade downflows (SADs), first observed on 1999 January 20 by the Soft X-ray Telescope (SXT) on \emph{Yohkoh} \citep{MK1999}, are dark voids that usually flow downwards above post-flare loops during the decay phase of the flare \citep{MK1999,2000SoPh..195..381M}. In the past decades, SADs were observed sporadically in soft X-ray channels \citep{MK1999,2000SoPh..195..381M,2001EP&S...53..581H,2010ApJ...722..329S}, white-light channels \citep{2002ApJ...579..874S}, and EUV channels \citep{2003SoPh..217..247I,2003SoPh..217..267I}. With the launch of \textit{Solar Dynamics Observatory} (\textit{SDO}; \citealt{Pesnell:2012aa}), SADs were observed more frequently by the higher temperature EUV channels like 131 \AA\ channel of Atmospheric Imaging Assembly (AIA; \citealt{Lemen:2012aa}) that is sensitive to the high-temperature plasma of about 10 MK (e.g., \citealt{2011ApJ...742...92W,2012ApJ...747L..40S,2014ApJ...796...27I,2014ApJ...786...95H,2016ApJ...829L..33L,2018ApJ...868..148L}). It is found that SADs may also appear during the impulsive phase, and are even correlated to hard X-ray bursts, indicating an important role of SADs in releasing flare energy \citep{Asai2004,khan2007}. 

With spectroscopic observations, \citet{2003SoPh..217..247I} for the first time concluded that SADs are often depleted regions, consistent with the suggestion proposed by \citet{MK1999}. This conclusion is further supported by the results based on the differential emission measure (DEM) technology (e.g.,  \citealt{2012ApJ...747L..40S,2014ApJ...786...95H,2017A...606A..84C}). The temperature of SADs is generally lower  than that of the surrounding structures \citep{2014ApJ...786...95H}. \citet{2017A...606A..84C} found that the EM of SADs above 4 MK is significantly smaller than that of the surrounding plasma. With a careful calculation, \citet{2017ApJ...836...55R} concluded that the adiabatic process could play an important role in heating SADs, as well as the surrounding plasma. The velocity of SADs is usually widely distributed, ranging from dozens to hundreds of kilometers per second, but smaller than the Alfv$\acute{e}$n velocity and the free-fall velocity \citep{MK1999,2000SoPh..195..381M,2009ApJ...697.1569M,2007ASPC..369..489M}. No matter what velocities the SADs have, they finally stop at the flare loop top, and the rate of deceleration, while uncertain, appears to be consistent with drag \citep{Linton2006,SheeleyJr2004,2010ApJ...722..329S,2011ApJ...730...98S,2011ApJ...742...92W,2016ApJ...829L..33L}. 

In the past decades, several models have been proposed to explain the formation of SADs. \citet{MK1999} proposed that SADs are the cross sections of evacuated flux tubes that are formed by magnetic reconnection in the wake of erupting coronal mass ejections (CMEs) (see also \citealt{2000SoPh..195..381M,2009ApJ...697.1569M}). \citet{2010ApJ...722..329S} pointed out that SADs and supra-arcade downflow loops (SADLs) are the same phenomena, and that the difference is only from their different viewing perspectives. In terms of intermittence of SADs, \citet{Linton2006} and \citet{2011ApJ...730...90G} suspected that the corresponding reconnection is actually patchy reconnection. \citet{2009EP&S...61..573L} supported the theory that SADs are reconnected magnetic flux loops because it is consistent with their rapid deceleration. Different from the interpretation of SADs directly from magnetic reconnection, \citet{2012ApJ...747L..40S} argued that SADs are wakes behind the retracting loops out of the reconnection region. This idea was confirmed to be a reasonable mechanism by the simulation results performed by \citet{2013ApJ...776...54S}. A different idea is that SADs are generated by expansion waves triggered by magnetic reconnection \citep{2009MNRAS.400L..85C,2010MNRAS.407L..89S,2011A&A...527L...5M,2012ApJ...759...79C,2015ApJ...807....6C,2016ApJ...832...74Z}. However, such a scenario predicted that the temperature of SADs is higher than that of the fan, which contradicts the observation results of \citet{2014ApJ...786...95H}. \citet{2013ApJ...775L..14C} proposed that SADs are channels carved by continuous reconnection outflow jets, which explains why SADs are not filled with the surrounding plasma immediately. \citet{2014ApJ...796L..29G} and \citet{2014ApJ...796...27I} suggested that SADs are a result of Rayleigh-Taylor instabilities between the reconnection outflow and the denser flare arcade fan. More recently, \citet{2020ApJ...898...88X} directly interpreted SADs as outflows of patchy and bursty reconnection as they can push away the surrounding plasma and the magnetic field keeps the plasma from flowing back.

In this paper, we statistically study 20 SAD events that took place in the decay phase of three flares with the aim mainly focusing on the thermal and kinetic evolutions of SADs, in particular the SAD front and body. We also study the statistical relationship between the pressure and temperature of SADs. The results are expected to be able to help understand the origin and dynamics of SADs. In Section \ref{data}, we introduce the data and methods. The results are presented in Section \ref{results}, which are followed by conclusions and discussions in Section \ref{discussion}.
%Data
\section{Data and Methods}
\label{data}

The AIA  on board \textit{SDO} observes the solar atmosphere in seven extreme ultraviolet (EUV) bands with a time cadence of 12 s, a spatial resolution of 0.6$^{\prime\prime}$ per pixel, and a field of view of 1.3 $R_{\odot}$. 
The AIA images with high resolution and high cadence enable to study the dynamics of SADs during the flare in detail. The AIA data are processed with the $aia\_prep.pro$ and deconvolved with $aia\_deconvolve\_richardsonlucy$ routines, both of which are available in the SolarSoft Ware (SSW) package \citep{Freeland:1998we}. We identify SADs with the AIA 131 \AA\ images as they are the most sensitive to the hot plasma emission of about 10 MK during the decay phase of the flare. We choose 20 SADs from three flares, which occurred on 2011 October 22, 2012 July 17 and 2015 June 18, respectively. Among these flares, the M1.3 flare on 2011 October 22 has been widely studied (e.g. \citealt{2012ApJ...747L..40S,2014ApJ...796...27I,2017ApJ...836...55R,2016PhDT........12S,2020ApJ...898...88X}).

First of all, we calculate the differential emission measure (DEM) of the flare regions. Among the seven AIA EUV channels, six channels, which are centered at 94 \AA, 131 \AA, 171 \AA, 193 \AA, 211 \AA\ and 335 \AA, are used. To increase the signal to noise ratio, the resolution of the EUV images are degraded to 2.4 arcsecs. The codes we use to calculate the DEM were proposed by \citet{2015ApJ...807..143C} and improved by \citet{2018ApJ...856L..17S}. Using the DEM, we further calculate the emission measure (EM) and DEM weighted average temperature defined as follows \citep{2012ApJ...761...62C}: 

\begin{equation}
\centering
EM = \int DEM (T) dT
\label{eq:em}
\end{equation}
\begin{equation}
\centering
T = \frac{\int T \times DEM(T) dT}{\int DEM(T) dT}
\label{eq:temp}
\end{equation}

\begin{deluxetable*}{c|ccc|ccc}
\tablenum{1}
\tablecaption{Properties of three flares and 20 SADs.\label{tab:para}}
\tablewidth{0pt}
\tablehead{
\multirow{2}*{Date} & \multirow{2}*{Start time} & \multirow{2}*{Peak time} & \multirow{2}*{Flare class}
& \multicolumn{3}{c}{SADs}\\
\cline{5-7}
 &  &  & & \colhead{ID} & \colhead{Start time} & \colhead{Duration (s)}
}
\startdata
\multirow{4}{*}{2011.10.22} & \multirow{4}{*}{10:00} & \multirow{4}{*}{11:10} & \multirow{4}{*}{M1.3} 
  & 1 & 12:02:21 & 336\\
 & &  &  & 2 & 12:14:33 & 204\\
 & &  & & 3 & 12:49:45 & 192\\
 & &  & & 4 & 13:01:45 & 312\\
 \cline{1-7}
\multirow{14}{*}{2012.07.17} & \multirow{14}{*}{12:03} & \multirow{14}{*}{17:15} & \multirow{14}{*}{M1.7}
  & 5 & 15:44:32 & 108\\
 & &  &   & 6 & 16:03:32 & 660\\
 & &  &   & 7 & 15:45:56 & 288\\
 & &  &   & 8 & 15:51:44 & 528\\
 & &  &   & 9 & 15:53:44 & 468\\
 & &  &   & 10 & 15:56:08 & 360\\
 & &  &   & 11 & 16:04:20 & 564\\
 & &  &  & 12 & 15:47:56 & 324\\
 & &  &  & 13 & 16:11:20 & 192\\
 & &  &  & 14 & 16:18:44 & 228\\
 & &  &  & 15 & 16:18:08 & 216\\
 & &  &  & 16 & 16:29:08 & 804\\
 & &  &  & 17 & 16:29:08 & 876\\
 & &  &   & 18 & 16:38:08 & 456\\
  \cline{1-7}
\multirow{2}{*}{2015.06.18} & \multirow{2}{*}{00:33} & \multirow{2}{*}{01:55} & \multirow{2}{*}{M1.2}
  & 19 & 01:48:32 & 132\\
 & &  &  & 20 & 02:00:20 & 240\\
\enddata 
\end{deluxetable*}

\begin{figure}[!ht]
\centering
\includegraphics[width=17cm]{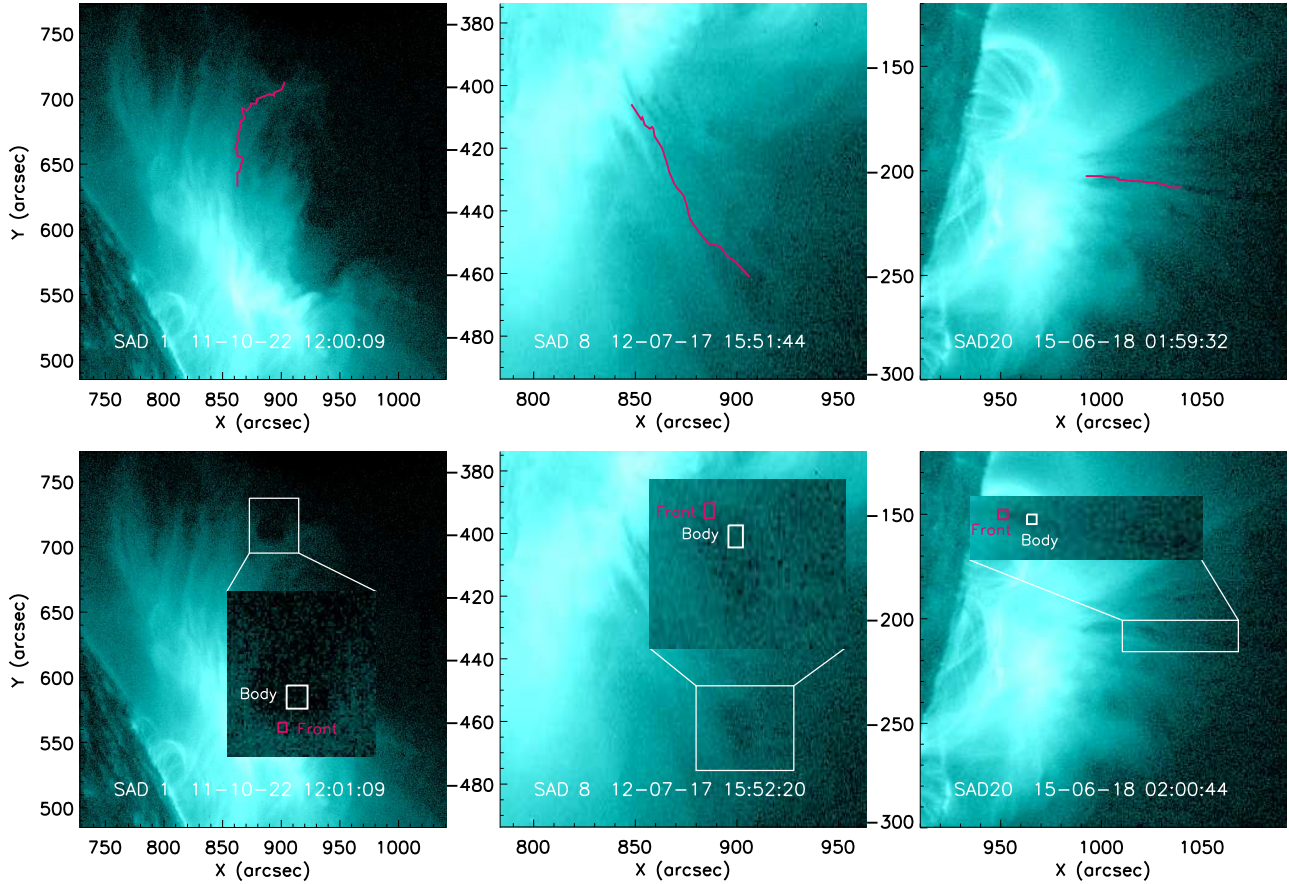}
\caption{First row: \textit{SDO}/AIA 131 \AA\ images showing the hot flare regions overlaid with paths of the front of SAD1, SAD8, and SAD20. Second row: Schematic diagram of the front position (red box) and the body position (white box).}
\label{fig:route}
\end{figure}

\begin{figure}[!ht]
\centering
\includegraphics[width=17cm]{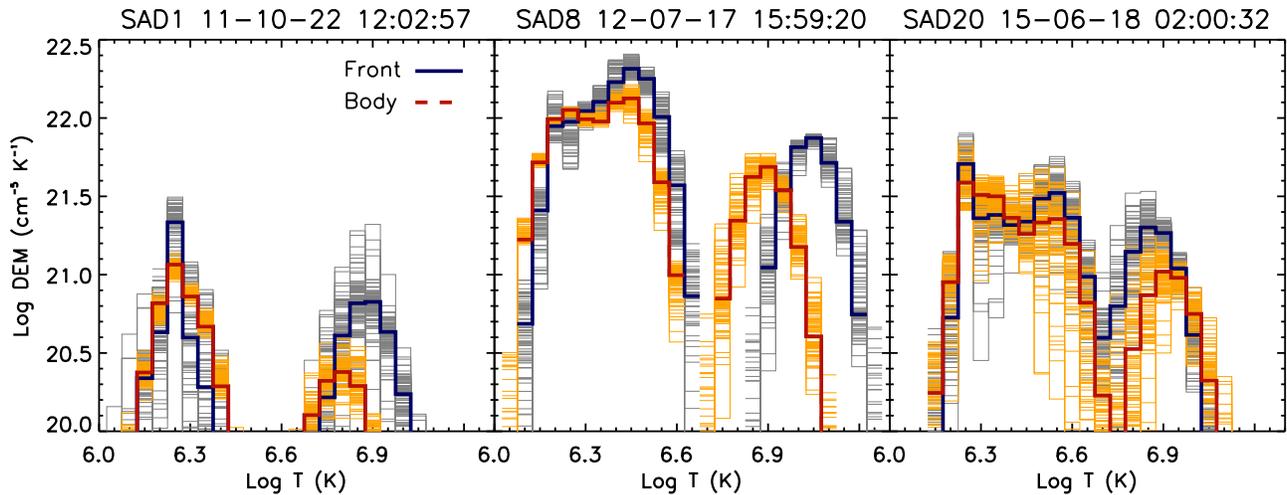}
\caption{DEM distributions for the front and body of three SADs (the same examples as Figure \ref{fig:route}). The blue lines represent the results for the front of SADs with gray lines showing the 100 Monte-Carlo simulations. The red lines represent the results for the body of SADs with orange lines showing the 100 Monte-Carlo simulations.}
\label{fig:dem_3}
\end{figure}

\begin{figure}[!ht]
\centering
\includegraphics[width=17cm]{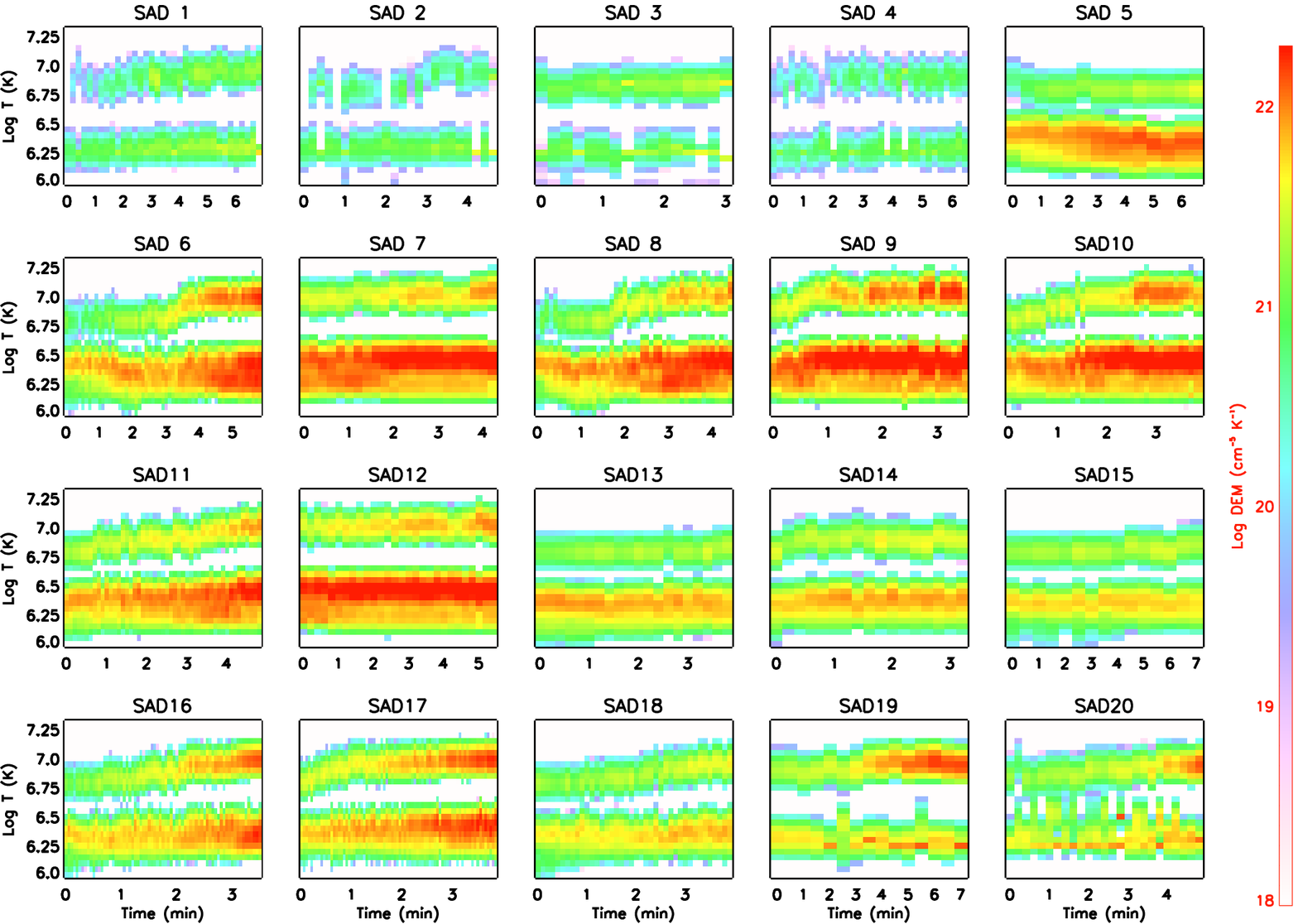}
\caption{Time evolutions of DEM distributions for the front region of 20 SADs. The DEM values are represented by different colors, as shown in the color bar on the right.}
\label{fig:front_dem}
\end{figure}

The DEM is integrated over the temperature range of log $T$ = 6.0 to log $T$ = 7.3.
To calculate the velocity of SADs on the plane of sky, we take advantage of Fourier Local Correlation Tracking (FLCT; \citealt{2008ASPC..383..373F}). The FLCT method is designed to estimate a 2D velocity field from two images with an assumption that the second image differs from the first one by the motion of features on the plane of sky. For each pixel, we calculate the cross-correlation coefficient between two images with one of them imposed by a displacement. Then the velocity for the pixel can be derived as the displacement that maximizes the cross-correlation coefficient divided by time resolution. Even though two-dimensional, FLCT velocities are regarded as the actual velocities of SADs because there seems to be no significant velocity component perpendicular to  the plane of sky when observed above the limb. We also use the cork-advection-method used by \citet{2013ApJ...766...39M} to examine the consistency. From the three flares, we select 20 SADs for our study. Their start time and duration are shown in Table \ref{tab:para}. The start time is defined as the moment when the SADs first appear in the 131 \AA\ images, and the end time is the time when they become invisible.

%Results
\section{Results}
\label{results}

%path
\subsection{Overview of SADs}
\label{path}

\begin{figure}[!ht]
\centering
\includegraphics[width=15cm]{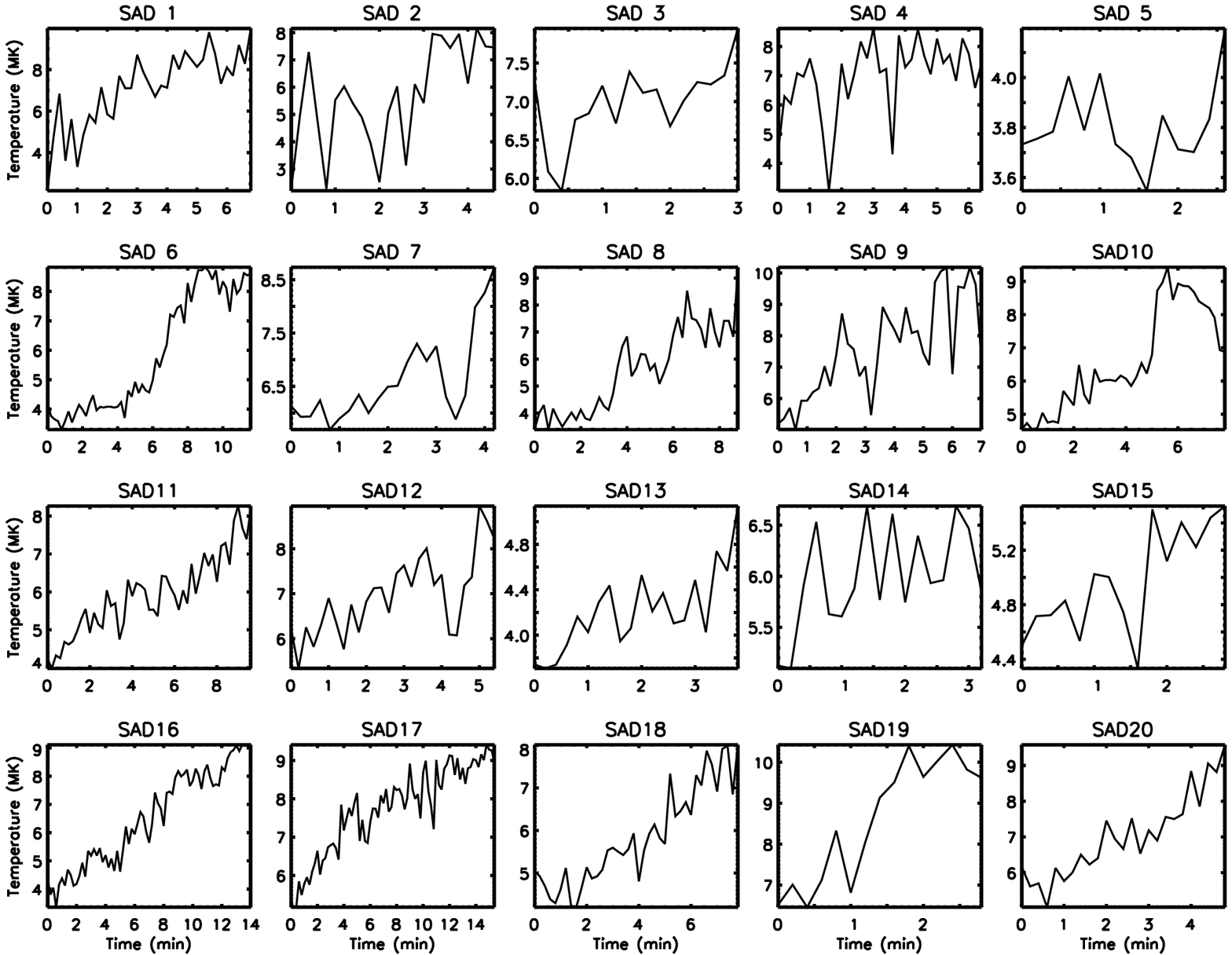}
\caption{Time evolutions of temperature for the fronts of 20 SADs.}
\label{fig:front_tem}
\end{figure}

\begin{figure}[!ht]
\centering
\includegraphics[width=17cm]{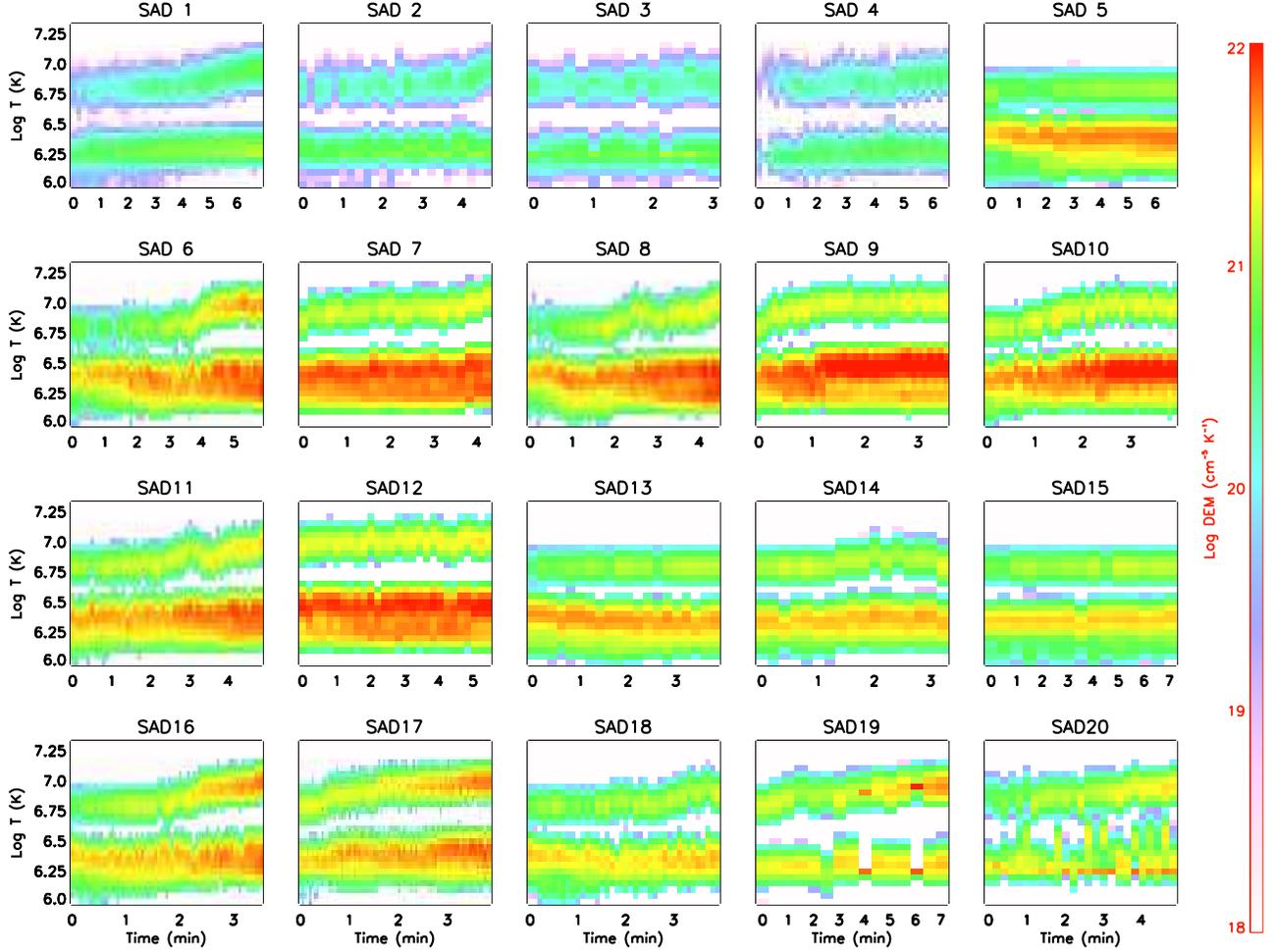}
\caption{Same as Figure \ref{fig:front_dem} but for the SAD body.}%
\label{fig:dimming_dem}
\end{figure}
\begin{figure}[!ht]
\centering
\includegraphics[width=15cm]{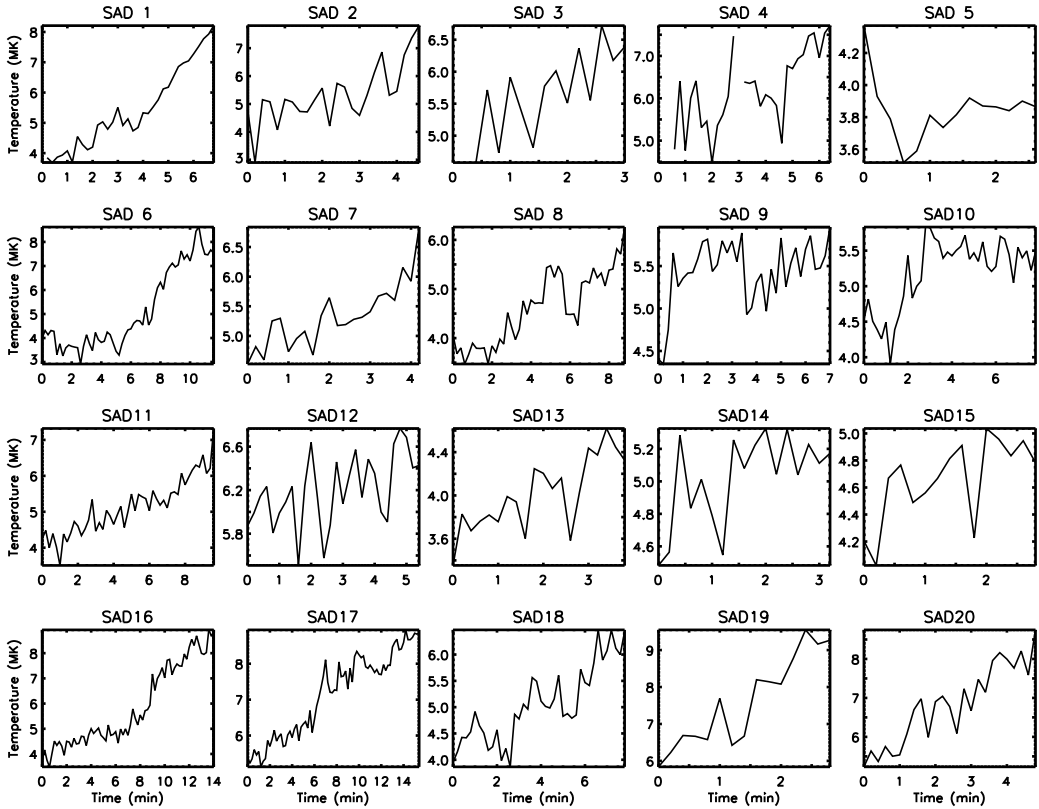}
\caption{Time evolutions of temperature for the bodies of 20 SADs.}
\label{fig:dimming_tem}
\end{figure}
\begin{figure}[!ht]
\centering
\includegraphics[width=15cm]{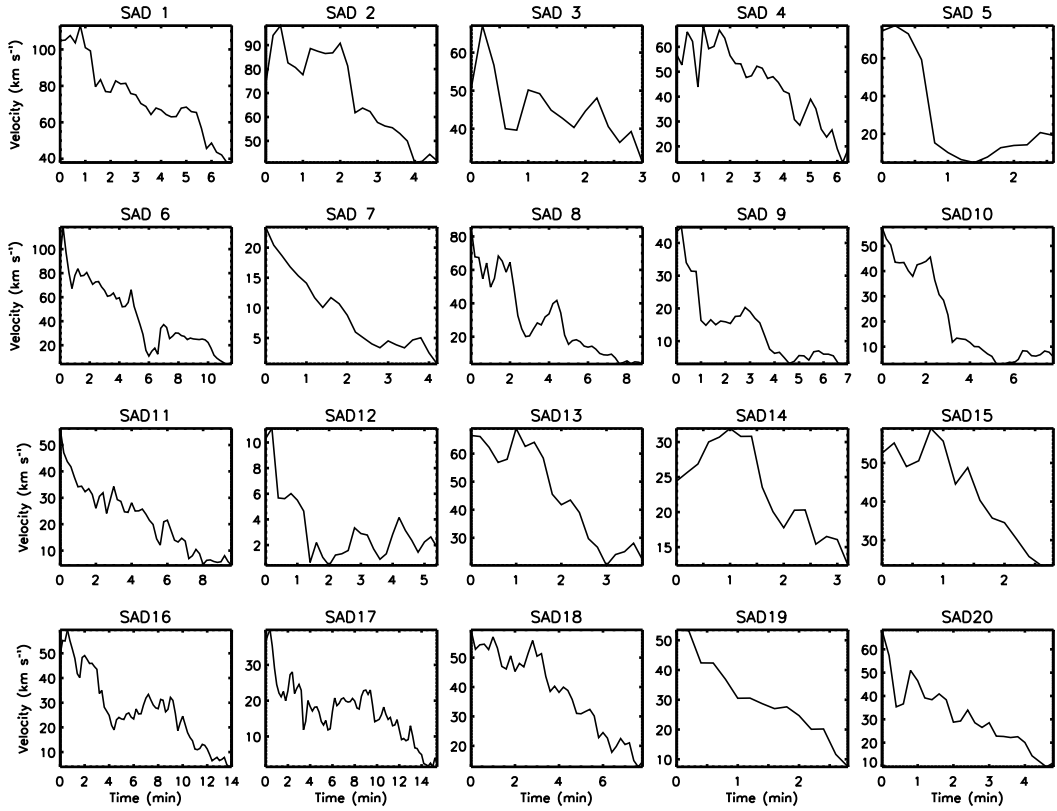}
\caption{Time evolutions of velocity for 20 SADs.}
\label{fig:front_v}
\end{figure}

\begin{figure}[!ht]
\centering
\includegraphics[width=15cm]{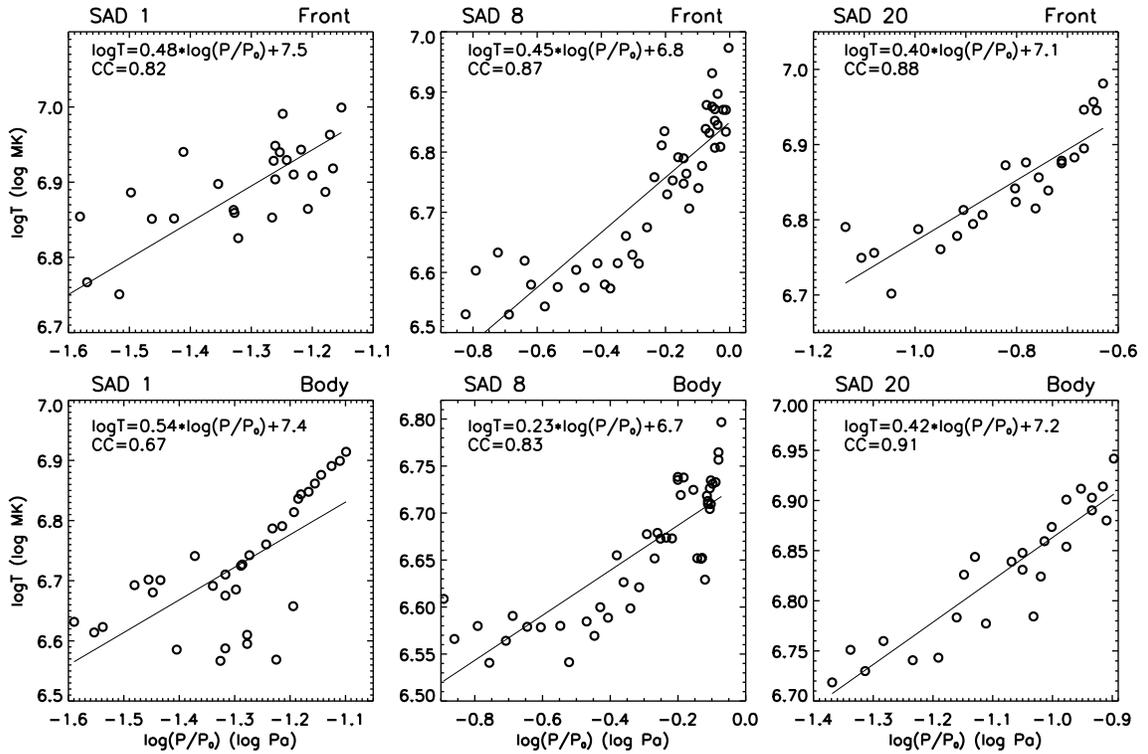}
\caption{Scatter plots of logarithmic temperature versus logarithmic pressure for the front and body of three selected SADs.}
\label{fig:pandt}
\end{figure}

\begin{figure}[!ht]
\centering
\includegraphics[width=15cm]{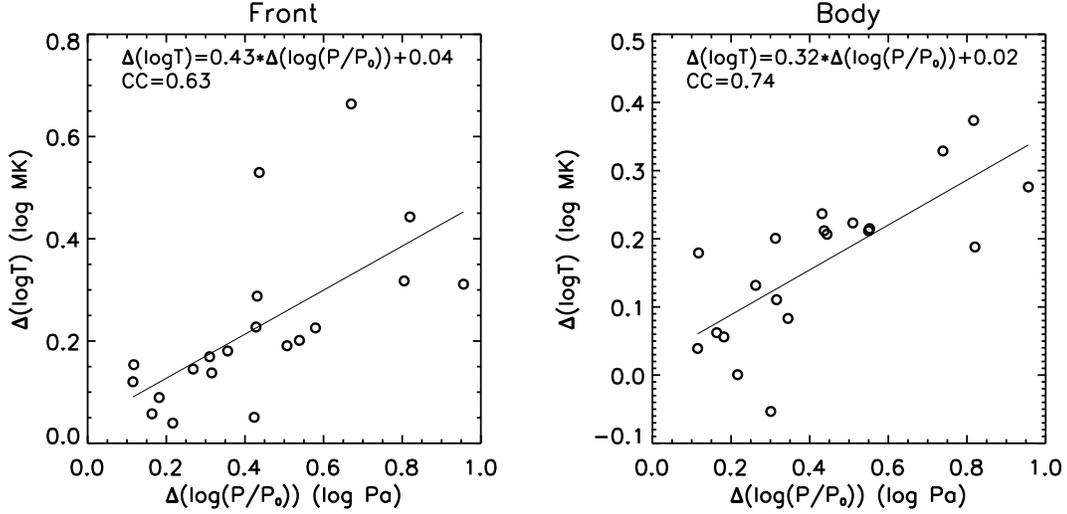}
\caption{Scatter plots of the variation of temperature versus that of pressure for the front (left) and body (right) of 20 SADs.}
\label{fig:p_delta}
\end{figure}

\begin{figure}[!ht]
\centering
\includegraphics[width=15cm]{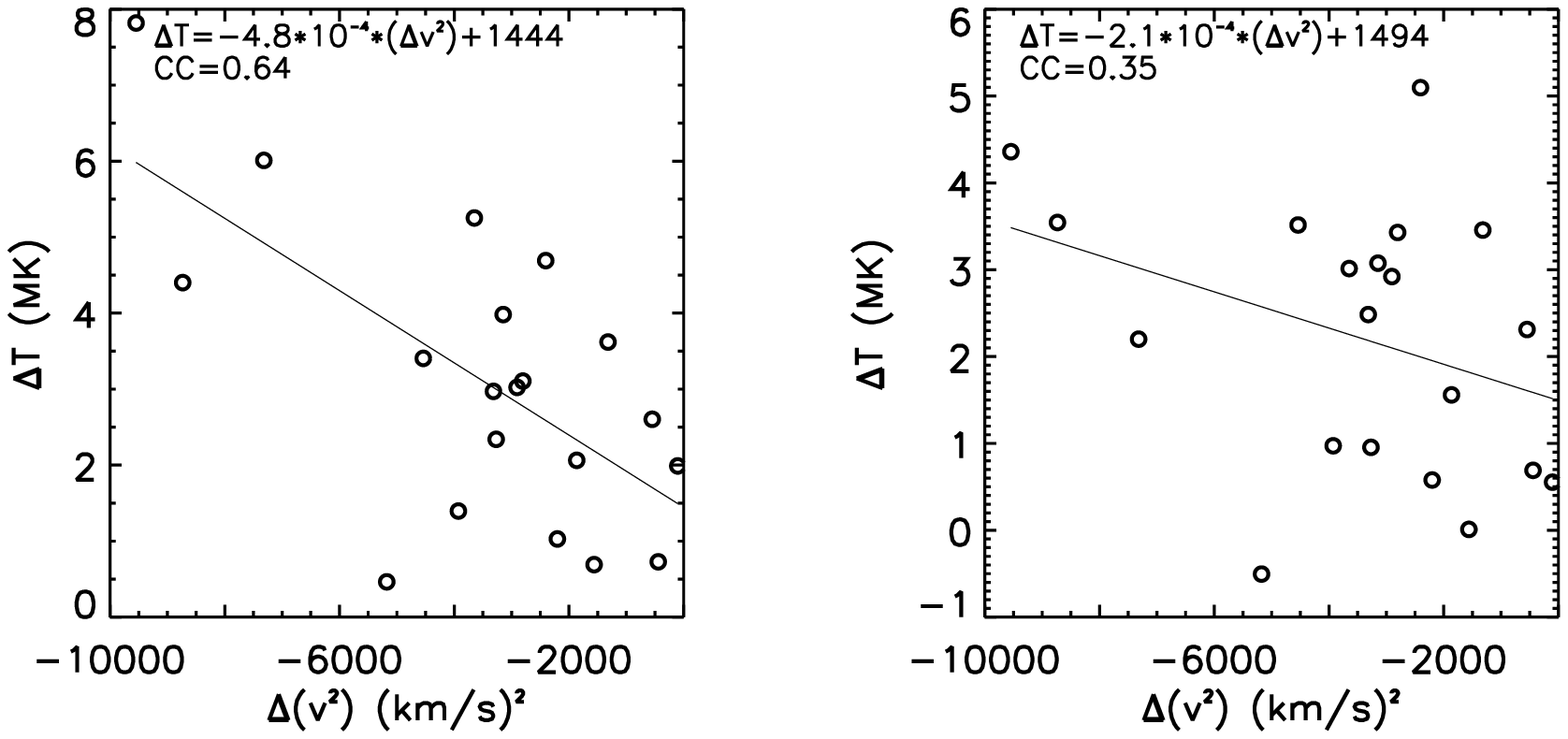}
\caption{Scatter plots of the variation of temperature versus that of velocity squared for the front (left) and body (right) of 20 SADs.}
\label{fig:correlation}
\end{figure}

SADs usually first appear as a dark structure at a high altitude above the solar surface. Afterwards, they quickly move downwards. Initially, they have a larger volume but rapidly become smaller after interacting with the hot plasma sheet of flares. Finally, they merge to the plasma sheet and become invisible, during which Kelvin–Helmholtz instability and turbulence may be involved (e.g., \citealt{2013ApJ...766...39M, 2018ApJ...866...64C}). For each SAD at each moment, the main body of SADs is identified as the dark void region in 131 \AA\ intensity images. The front of SADs is the region precedes the void in its flow path.
For each SAD, we select two square boxes as the proxies of the SAD front and body (examples are shown in the second row in Figure \ref{fig:route}). The size of the box representing the SAD front is about 2 pixels $\times$ 2 pixels and does not change for different SAD events. However, for the SAD void, we have to adjust the box size for the sake of including the void region as much as possible for each SAD. We then mark the front in each frame and connect them to get the path of the SAD. It represents the trajectory of the SAD. The trajectories for three selected SADs (SAD1, SAD8 and SAD20) are shown as red lines in Figure \ref{fig:route}.

%thermal
\subsection{Thermal Properties}
The DEM distributions of the SAD front and body at three selected moments are shown in Figure \ref{fig:dem_3}. The blue and red solid curves indicate the best DEM solutions to the observed AIA intensity for the front and the body, respectively. The uncertainties of DEM solutions are estimated using 100 Monte-Carlo simulations. At each temperature bin, the DEM solution is considered to be reliable if the dispersion of 100 MC solutions is small and converges toward best solution.  Obviously, the DEM solutions are well resolved in the temperature range from log $T$ = 6.0 to log $T$ = 7.3. The evolutions of the DEM for the fronts of 20 SADs are displayed in Figure \ref{fig:front_dem}. One can clearly see that all DEM distributions have two peaks, one at a low temperature of about 2 MK, the other at a high temperature of about 7 MK - 10 MK. The low temperature component barely changes with the evolution of SADs, because it comes mostly from the background coronal plasma. The high temperature component is mainly from the emission of the SADs themselves. Both the intensity we have observed and the DEM we have derived are always from an integration along the line of sight. Thus, it is not possible to isolate the contribution of SADs from that of the foreground and background. For some cases, like SAD12, although the EM of the high temperature component significantly increase, the peak temperature has almost no change. While for some other cases, like SAD6 and SAD8, the peak temperatures are shifted towards higher temperature values, e.g., from about log $T$ = 6.8 to log $T$ = 7.0. These results show that the SAD fronts are heated as the SADs fall down.

The heating of the SAD fronts is further confirmed by the evolution of DEM weighted temperature as shown in Figure \ref{fig:front_tem}. In particular, even for those cases (like SAD3 and SAD 13) whose EM of the high temperature component does not increase, the average temperature still increases slightly. For most cases, the average temperature increases by more than  30\%. In all the cases, the average temperature reaches nearly up to 10 MK when the SADs stop at the flare loop top.

Similarly, we also display the DEM distributions of the SAD main body in Figure \ref{fig:dimming_dem} and the evolution of their average temperature in Figure \ref{fig:dimming_tem}. We find that, similar to the SAD front, the EM of the high temperature component for the main body gradually increases as the SADs move downwards. The downward motion also results in an increase of the average temperature. However, it is worth noticing that although the SAD front and body show similar properties in thermodynamics, there is still a quantitative difference, i.e., the proportion of the high temperature component relative to the total EM for the SAD front is higher than that for the SAD body. In most cases, the peak temperature in the DEM distribution for the front is obviously higher than that for the body, as clearly revealed in some cases like SAD1 and SAD8, as illustrated in Figure \ref{fig:dem_3}. It indicates that the temperature of the SAD front is higher than that of the body, which agrees with  the conclusion of \citet{2014ApJ...786...95H}. For SAD20, even though the peak temperature (log $T$ = 6.85) in the hotter component for the front is a little lower than that (log $T$ = 6.90) for the body, the average temperature of the front is still higher, which is mainly due to a larger volume of high temperature plasma in the front.

%dynamic
\subsection{Dynamics of SADs}
Using the FLCT method, we can derive the velocity of the SAD front, which can be approximately regarded as that of the SAD body. Figure \ref{fig:front_v} shows the velocity evolutions of the 20 SADs. One can see that all SADs gradually decelerate during the course of motion toward flare loops. The initial velocities of SADs are of the order of 100 km s$^{-1}$ that decrease to almost zero in minutes. The average deceleration is about 0.15 km s$^{-2}$. 

%relationship
%\subsection{Analysis of Heating Mechanism}
It has been suggested that the heating of the SAD front may be due to adiabatic compression \citep{2017ApJ...836...55R}. Here, we further explore this issue statistically.
For an ideal gas, an adiabatic process means that the temperature and pressure obey the following equation, 
\begin{equation}
\centering
\frac{P^{\gamma-1}}{T^{\gamma}} = C
\label{eq:adiabatic}
\end{equation}
where $\gamma = 5/3$ is the polytropic index and $C$ is a constant. We notice that the deceleration of SADs is relatively small and thus assume that the atmosphere above flare loops is in static equilibrium, the pressure decays exponentially with height as:
\begin{equation}
\centering
P = P_0{\rm exp}(-H/H_0)
\label{eq:pressure}
\end{equation}
where $P_0$ is a dimensionless constant, $H_0$ = 30 Mm is the average scale height at $T$ = 1 MK. The heights of the front and body of SADs at each moment are measured relative to the solar surface. 

In Figure \ref{fig:pandt}, we present the scatter plots of temperature vs. pressure in logarithmic scale for the front and body of three selected SADs. It is found that the two parameters are related linearly, the correlation coefficients is as high as 0.8. The linear-fitting between log $T$ and log $P$ yields a coefficient close to 0.4 (e.g., $P \sim T^{5/2}$), suggesting that adiabatic compression plays an important role in heating the plasma at the front and in the body of SADs. 

We further study the relationship between the variation of pressure and that of temperature for the SAD front and body. The parameters we calculate are $\Delta$ (log$T$) and $\Delta$ (log($P/P_0$)). For each SAD, the variations are estimated by subtracting the value at the first moment from that at the end moment. Figure \ref{fig:p_delta} shows the statistical result. The correlation coefficients for the two parameters are about 0.63 and 0.74 for the SAD front and body, respectively, which further supports the interpretation that the adiabatic heating plays an important role in heating the different structural components of SADs. 

In order to check the possibility that the kinetic energy of SADs may be converted into thermal energy through viscous dissipation, we further investigate the statistical relationship between $\Delta (V^2)$ and $\Delta T$, which are proxies of the variations of kinetic energy and thermal energy, respectively. For the 20 SADs we study, the result is shown in Figure \ref{fig:correlation}. One can find that, for the SAD front, the increase of temperature seems to be moderately related to the decrease of kinetic energy, with a correlation coefficient being about 0.64; while for the SAD body, the increase of temperature is only weakly related to the decrease of kinetic energy, with a low correlation coefficient of about 0.35.
This implies that the dissipation of the kinetic energy of SADs may also play a role on the heating at the fronts but not the bodies. This also suggests that the heating occurs mostly at the front region that is not conducted to the body of the SADs. However, through further calculating the values of the kinetic energy and the thermal energy, we find that the kinetic energy is ten times smaller than the thermal energy both for the SAD front and body. This shows that the role of the viscous dissipation in heating the SAD front is limited as compared to the adiabatic process.
 
 %Discussion
\section{Summary and Discussions}
\label{discussion}
 %%%%
 In this work, we analyze the thermodynamic properties of 20 SADs. We show that the EM of the SAD main body at the high temperature range is smaller than that of the front region (Figure \ref{fig:dem_3}), which is consistent with previous results that SADs have a smaller EM than the surroundings (e.g., \citealt{2012ApJ...747L..40S,2014ApJ...786...95H,2017A...606A..84C}). However, it is noticed that the surrounding regions in previous studies were randomly chosen, which may bring about a significant uncertainty in comparisons. In our work, we select the front region, which can be determined precisely. Our approach is similar to that of \citet{2017ApJ...836...55R} but including more cases from different flares. We find that the temperature of the SAD front increases when the SAD falls down, which is consistent with the results of \citet{2017ApJ...836...55R}. The similar result obtained for different cases indicates that it could be a common phenomenon. The temperature of the SAD body evolves similarly. These results suggest that SADs are heated up when moving downwards. On the other hand, \citet{2020ApJ...898...88X} analyzed the temperature evolutions of fixed regions and also found that the temperature there increases after the SADs have passed by. 
 
Besides, we compare the front of SADs with the SAD main body. We find that the front of SADs is usually hotter than the following main body. It is different from the models developed by \citet{2014ApJ...796L..29G}, \citet{2014ApJ...796...27I}, and \citet{2015ApJ...807....6C}, which predict a higher temperature in the SAD body. The main reason could be that these models neglect thermal conduction and thus overestimate the temperature of the SAD body. We should also point out that not all SADs are cooler than the surroundings. For example, among the SADs examined by \citet{2014ApJ...786...95H}, one SAD is found to be slightly hotter than the surrounding.

In our work, we for the first time show the DEM evolution of different structure of SADs. The DEM temporal variations clearly indicate the increases of the peak temperature and the EM values. This greatly replenishes previous studies, in which the DEM distributions were only derived at some selected moments (e.g., \citealt{2012ApJ...747L..40S,2014ApJ...786...95H,2016ApJ...829L..33L,2017A...606A..84C,2017ApJ...836...55R,2020ApJ...898...88X}).

The most important result derived here is the good correlation between the pressure and  temperature in logarithmic scale for both the front and main body of SADs. The linear fitting between logarithmic temperature and logarithmic pressure yields a coefficient very close to 0.4. After examining all SADs, we find that this coefficient is distributed from 0.23 to 0.86 with the average values being 0.44 and 0.41 for the SAD front and body, respectively. It suggests that both the front and body of SADs may evolve following an adiabatic process and the adiabatic compression could be the critical mechanism of heating SADs when they move downwards, which confirms the conclusions in \citet{2017ApJ...836...55R} from, however, a more detailed statistical basis.

Note that, the background atmosphere is assumed to be in static equilibrium in our calculations. This assumption may be not valid for the dynamic flare loops, but may be a reasonable assumption for the atmosphere above the flare loops, where the SADs flow downward. In our assumption, we take $P_0$ to be 3.1 Pa according to Table 14.2 in \citet{2000asqu.book.....C}. Thus, the pressure we calculate under this assumption ranges from $0.30$ to $29\ dyne\ cm^{-2}$ for the atmosphere the SADs flowing through (from 140 Mm to 2 Mm above the solar surface). In order to test the reliability of such an assumption, we also directly calculate the gas pressure by equation $P = NkT$, where $N$ is the total number density of plasma  and $k$ is the Boltzman constant. By assuming a fully ionized hydrogen atmosphere, $N=2n_{e}$. The electron number density is calculated by $n_{e} = \sqrt{EM/l}$ where $l = 10^9$ cm is the average depth of the SADs along line of sight. Thus, the value of $n_e$ lies in a range of $3 \times 10^9$ to $1 \times 10^{10}$ $cm^{-3}$. We finally get a pressure that is approximately in a range from $0.23$ to $16\ dyne\ cm^{-2}$. Therefore, the calculated pressure is very close to the pressure estimated through the static atmosphere assumption, which validates our method.

We further compare the correlation between the variation of temperature and that of kinetic energy. We find that, for the SAD front, the increase of temperature is related to the decrease of kinetic energy. However, for the SAD body, the correlation does not hold or is very weak. Considering that the kinetic energy is only one order of magnitude lower than the thermal energy of the SADs, we suggest that the viscous dissipation may play a very limited role in heating the SAD front although its total kinetic energy is believed to be totally dissipated to thermal energy \citep{Linton2006}. On the other hand, \citet{2017ApJ...836...55R} analyzed quantitatively the various heating sources, including adiabatic and viscous processes, as well as conductive cooling. They also concluded that the adiabatic process is the main heating mechanism of the plasma at the font of SADs, while viscous process can almost be ignored. A recent study by \citet{2020ApJ...898...88X} also found that heating could be explained by adiabatic compression. Our results agrees with their conclusions that the heating of SADs is mostly from the adiabatic compression while the contribution of viscous dissipation is limited.

Note that the velocity we have derived is only the projected component of the three-dimensional velocity in the plane of sky; thus the kinetic energy of SADs we have estimated is only a lower limit. Regardless of the uncertainty caused by the projection effect, the initial velocities and their temporal evolutions of SADs are similar to previous results (e.g., \citealt{MK1999,2000SoPh..195..381M,SheeleyJr2004,Linton2006,2009ApJ...697.1569M,2007ASPC..369..489M,2010ApJ...722..329S,2011ApJ...730...98S,2011ApJ...742...92W,2016ApJ...829L..33L}). The fact that the initial velocities are lower than the local Alfv$\acute{e}$n velocity and the free-fall velocity cast doubt on the interpretation that SADs directly correspond to the reconnection outflows. The deceleration of SADs could be mainly attributed to the drag force, a result of the interaction of SADs with the background flare sheet. However, the model of SADs created by Rayleigh-Taylor instabilities proposed by \citet{2014ApJ...796L..29G} may explain the much lower velocity of SADs. They found that the velocity of SADs is the growth rate of the instabilities, which is about 5\% of the Alfv$\acute{e}$n velocity.

\acknowledgments
We thank the referee who helped improve the readability and clarity of the paper. We also thank the team of SDO/AIA for providing the valuable data. AIA data are courtesy of NASA/SDO, which is a mission of NASA’s Living With a Star Program. Z.F.L., X.C., and M.D.D. are funded by NSFC grants 11722325, 11733003, 11790303 and 11790300. K.K.R. is supported by NSF SHINE grant AGS-1723425 and NASA grant 80NSSC18K0732.  D.K. is supported by the NSF-REU solar physics program at SAO, grant AGS-1850750. M.W. and D.M. are supported by NASA grant 80NSSC18K0732.

\bibliographystyle{aasjournal}

%\listofchanges

\end{document}